# Study of plume dynamics and plasma density during its expansion

*G. Veda Prakash, Kiran Patel, Narayan Behera, and Ajai Kumar*
*Institute for Plasma Research, Bhat, Gandhinagar, Gujarat, HBNI, Mumbai*
*prakashgveda@gmail.com

The electron density is one of the crucial parameters in plasma plume, which plays a major role in producing the key reactive species required for biological and industrial applications. In this work, experiments are performed to estimate the spatial variation of plasma density in a steady plume. The time-resolved images obtained from the ICCD camera, plume current measurements are utilized for estimating the plasma density. We observed the plasma density increases from the glass nozzle to the plume tip. This increase in plasma density is due to the combined effect of reduction of drift velocity, charge, and area confinement of plume. Further, the intensity plots of the ICCD camera images obtained with larger exposure time validate the increased density at the plume tip. The plasma density along the plume gives information about the approximation to the plume exposure time desired and the impact it creates on the very sensitive targets. The spatial variation of plasma density in a plume is a futuristic requirement for several biological and industrial applications considering a single plasma device.

Over the last few years, low-temperature atmospheric pressure plasma is one of the fastest growing areas for fundamental study and due to its promising applications in biological and industrial areas such as wound healing, tissue sterilization, tissue engineering, surface modification, the sterilization of heat sensitive materials and instruments, and textile industry, etc.[1-3]. The plasma generated by using gases such as He, Argon, air, etc. are sources of reactive species, which can take part in various chemical reactions. In order to use this plasma for biological and industrial applications, it is vitally necessary to thoroughly understand the plasma parameters such as electron density ($n_e$), electron temperature ($T_e$), gas temperature ($T_g$), active species content, etc. In this direction, several experimental studies have taken place in the past to understand the above plasma parameters using various diagnostics such as electrical diagnostics, optical emission spectra (OES) [4-5] etc. In the above, the electron density is one of the crucial parameter, which yields the reactive species required for therapeutic applications. Several authors have estimated the electron density by using optical emission spectroscopy (OES) technique [6-8], and charge density method [9] and it is found to be in the order of ~ $10^{12}$ cm$^{-3}$. An overview of the range of electron densities that are encountered in typical atmospheric pressure plasmas is described in.[10] The utilization of OES for estimating the plasma plume parameters will include a great amount of signal to noise ratio, and which may lead to overestimation of the plasma parameters. In addition, the utilizing the microwave interferometry technique is much expensive and involve complexity in the analysis of post-experimental data. In this work, for the very first time, we have demonstrated a method to estimate the plume density along the especially along the plume length taking into consideration of in-situ information, low complexity in the data analysis, and cost-effective method.

In this work, the spatial variation of plasma density estimation along the plume length is carried out by taking into consideration of a wide range of biological and industrial applications where, target is sensitive to the applied plasma shape (area, pressure, etc.) and plasma parameters (gas temperature, electron density, electron temperature, etc.).[11-16] The plume dynamics such as the plasma propagation, geometrical structural variation of the flowing plasma, temporal evolution into the ambient air are estimated using ICCD images captured along the plume length. We have used a drag force model to understand the plume interaction with the ambient atmospheric air. The drift velocity (obtained from time-resolved images) and plasma current (measured using current transfer) measurements are utilized to estimate the plasma density along the plume length.

In this work, the cross-field geometry is utilized with glass tube inner diameter 4 mm and thickness 1 mm. Helium (He) gas of 99.99% purity is used as an active gas. The photograph of the visible steady plasma plume formed in the ambient air at a discharge voltage of 4 kV p-p and frequency of 33 kHz [10] with He flow rate of 11 lpm is shown in Figure 1.

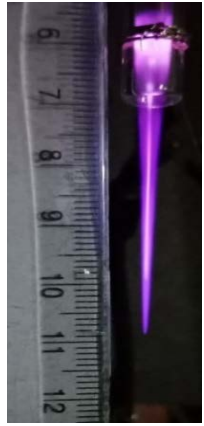

Figure 1 (Colour online) Photograph of the visible plasma jet with 4 kV$_{p-p}$ applied voltage, 33 kHz frequency, and gas flow rate 11 lpm.

In this work, our objective is to estimate the plasma density along the plume length. In a weakly ionized plasma, the relation between the current density is proportional to j = -n$_e$ev$_{drift}$, [17]. In order to determine the plasma density, the above relation can be further simplified as

$$n_{plume} = \frac{I_{plume}}{ev_{drift}S} \qquad (1)$$

where *e* is the elementary charge (*e* = 1.609× 10$^{-19}$ C), *I$_{plume}$* is plasma plume current, *v$_{drift}$* is plume drift velocity. So the spatial variation of plasma density can be estimated using the plume velocity, plume area and plasma current at particular location by using equation 1. *v$_{dirft}$* and *I$_{plume}$* are obtained from the ICCD camera images and current transformer (CT) respectively. For this, Time-Resolved fast images (500 ns dealy time) of flowing plasma plume are captured using an intensified charge coupled device (ICCD, 4 Picos, Stanford computer optics) camera fitted with f/1.4 apparatus (AVENIR lens) 25 mm lens and current transformer (FCT-016-20:1, Bergoz instrumentations, 1.25V/A) is used for measurement of plasma plume current (*I$_{Plume}$*).

The time-resolved images of visible plasma plume provide detailed information related to the hydrodynamic and geometrical structure of flowing plasma plume. This temporal evolution of plume has been acquired using ICCD (4 Picos, Stanford Computer Optics) camera by varying the delay time from 500-6000ns with reference to zero crossing voltage and exposure time of 500ns. The camera and the plasma voltage source synchronized using oscilloscope auxiliary output. At the instant of voltage signal been captured by an oscilloscope, it delivers a pulse output voltage from the auxiliary output with reference to the trigger point in the oscilloscope. The trigger reference is taken from the leading edge of the voltage pulse. A delay time of 500 ns increased for succeeding images of the plume. For better visibility, grey images have been converted into pseudo-color images using jet-color map in MATLAB. Figure 2 shows time-resolved ICCD images of the plasma jet emitted from the cylindrical glass tube illustrating individual plasma bullets. To estimate the velocity of the plasma plume, a mesh image of known dimensions is recorded to map the physical dimension of the plasma plume under same camera settings. The repeatability of plasma bullets has been confirmed by recording three images under similar experimental conditions. Size of the plume is estimated by a segmentation algorithm using MATLAB.[18] Plasma plume generated with the kHz frequency and a few kV produce a long plasma jet, which seems continuous to the naked eye, but on microsecond timescales, it is a fast-moving plasma packet or bullets. The diameter of the plume at the nozzle is observed approximately equal to inner diameter of the glass tube. This sub-microsecond time-resolved information is also essential for theoretical modeling and simulation work similar like laser produced plasma plume.[18]

It is observed from the images that, the initial velocity (i.e. near the glass nozzle) is around 18.6 km/s. After traveling ~ ¾ th distance of the plume, the velocity becomes 2.1 km/s and is remain constant till the end of the plume tip as shown in Figure 3. At the initial, the propagation velocity of the plume depends on the charge and electric field present at that moment. As the plume passes through the ambient air, the charges present in the plume are neutralized and a net decrease in charge or field experiences the opposition or drag that ultimately reduces velocity at the front of the plume. In addition to this, the plume also experiences the drag force proportional to its velocity from the viscous ambient air. So the equation of motion, acceleration a = - βv, which gives the velocity v = $v_o$ exp (- βt) and the plume position z = $z_f$ [1 – exp (−βt)] [19,20]. This model predicts that the plume will eventually come to rest, as a result of resistance from the collisions with the background gas. The length of the plasma plume *vs* delay time curve is fitted with drag force model z = $z_f$ [1 – exp (−βt)][20]. The best-fitted parameters i.e. slowing co-efficient β ~ 0.4 µs$^{-1}$ and stopping distance $z_f$ ( = $v_o$/ β) ~ 36.1 mm. Throughout the plume propagation, it has two velocities i.e. an initial fast moving plume travelling at 18.6 km/s and a reduced velocity 2.1 km/s due to drag force. This model predicts the information of plume velocity and the propagation distance of the plume in the ambient air. This best fitting of drag model confirms that in addition to the electric field, the ambiance drag also affects the plume velocity.

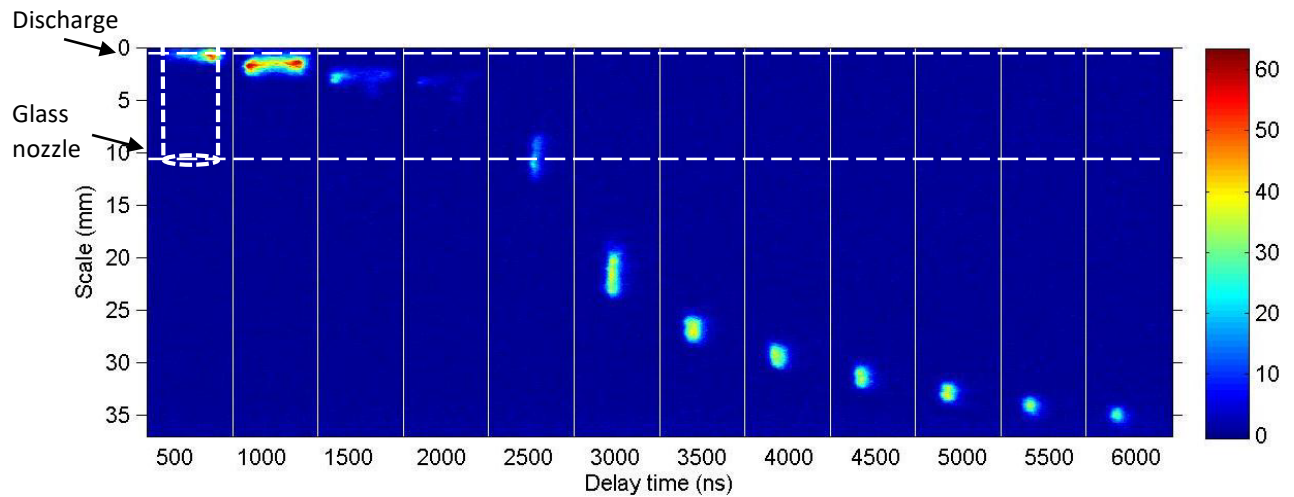

Figure 2 (Colour online) Sequence of time resolved ICCD images of the flowing plasma bullets propagation into ambient air at 11 lpm gas flow rate.

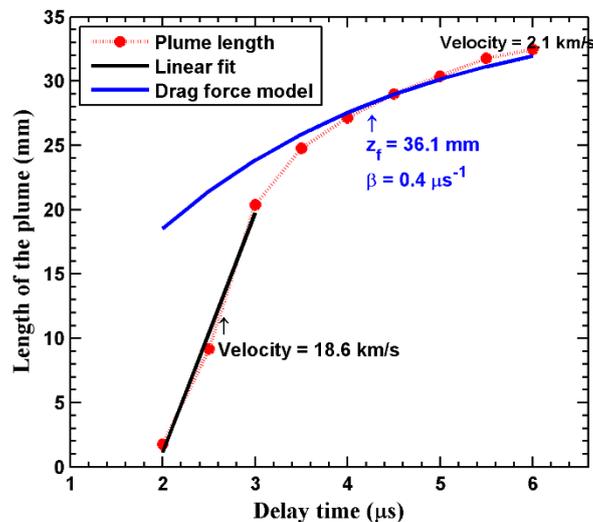

Figure 3 (Colour online) Time evolution of Plasma plume, dashed red dotted bullet line represent the experimental data, Solid black and solid blue line represent the linear fit and fitting for drag force model.

In the atmospheric-pressure plasma plume, due to lighter in weight, an electron accelerates by the electric field whereas ion hardly moves; hence, the electron attachment is normally ignored. In addition, this plasma bullets propagation phenomenon is similar to a positive streamer caused by photoionization. In photoionization, the ionization front travels at a velocity qualitatively similar to the drift velocity of the electron, as the movement of the heavy ions is almost negligible. Therefore, under the atmospheric pressure, the drift velocity of the charge carrier is qualitatively equal to the drift velocity of the plasma bullet.[9, 21-22] In the present case, the measured drift velocity from glass nozzle to the plume tip is in the range of 18.6 - 2.1 km/s.

The plume current is estimated as a charge flow along it. In this work, direct measurement of plume current has been carried out at three different locations along the plume length.[5]. The plasma plume current amplitude changes with time and distance of the plume travel length. The CT has inner diameter of 10 mm, and the plume has a maximum diameter of 3.3 mm. Hence, the plume never affected with the arrangement of CT along the length of the plume. The variation of plasma plume current along the length is shown in Figure 4.

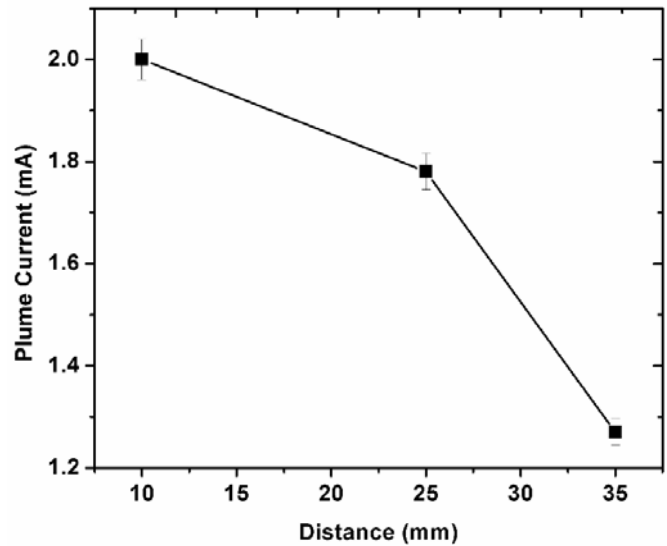

Figure 4 Variation of plasma plume current along the length of the plume.

By using the plume velocity and current information, the plume density is calculated. The plasma density for three different positions of the plume i.e., at 10, 25, and 35 mm respectively from the main discharge is 0.05879, 0.13068, and $3.26 \times 10^{12}$ cm$^{-3}$ respectively which are in the limits of earlier reported values.[23]. Here, it can be observed that the plasma density values increase from the glass orifice to the plasma plume tip. In this, it should be noted that the variables on the right-hand side of the equation.1 ($I_{plume}$, $v_{drift}$, and S) are decreasing from the glass orifice to the plasma plume tip. In order to explain the increasing trend of the $n_e$, $I_{plume}$ (numerator of the equation.1) is normalized with the $v_{drift}$ .S (denominator of the equation.1) and it is denoted as $I_{plume, norm}$. The rate of decrement in $v_{drift}$ .S found to be much higher compared to that of $I_{plume,norm}$ and hence the $n_e$ increases as shown in Figure 5. Further, the precise variation of density can be obtained by conducting more theoretical work from PIC simulation[24] which in not the scope of present work.

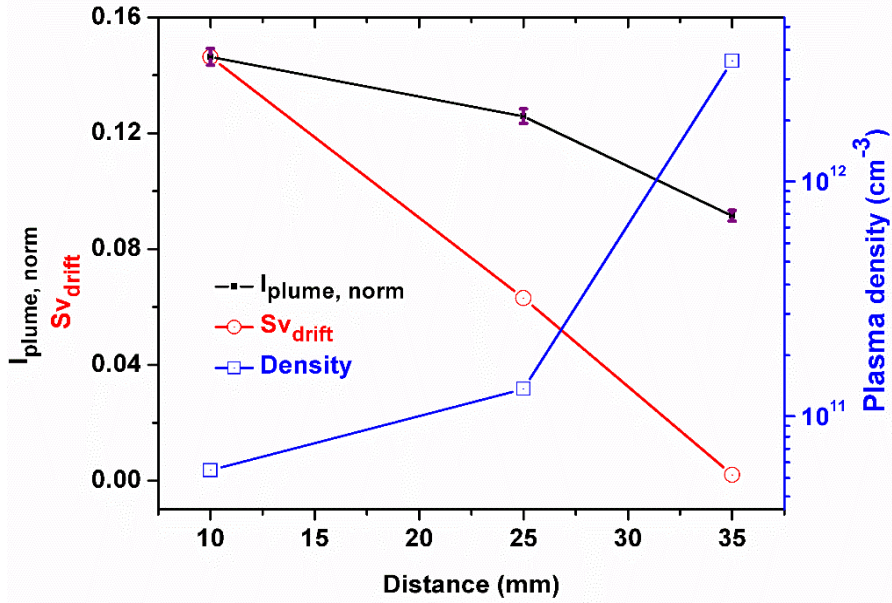

Figure 5 (Colour online) Variation of $I_{plume,norm}$, $Sv_{drift}$, and density (± 2% error) along the plume length.

Further, the increase in the plasma density at the plume tip is validated by using digital image processing as shown in Figure *6*. Figure 6 (a) shows the pseudo-colour image of ICCD camera image of flowing plasma with 20 us exposure time. Figure *6* (b) shows the intensity profile of the plume over the length. The increase of intensity profile shows the line integrated photon counts, which is proportional to the plasma density.

This study will helpful for futuristic fundamental research related to the spatial variation of plasma density along the length of the plasma. These plasma parameters i.e. velocity and plasma density information along the plume length helps to decide the duration of the treatment and also helps to estimate the impact due to high electron density at the plume tip for very sensitive targets.

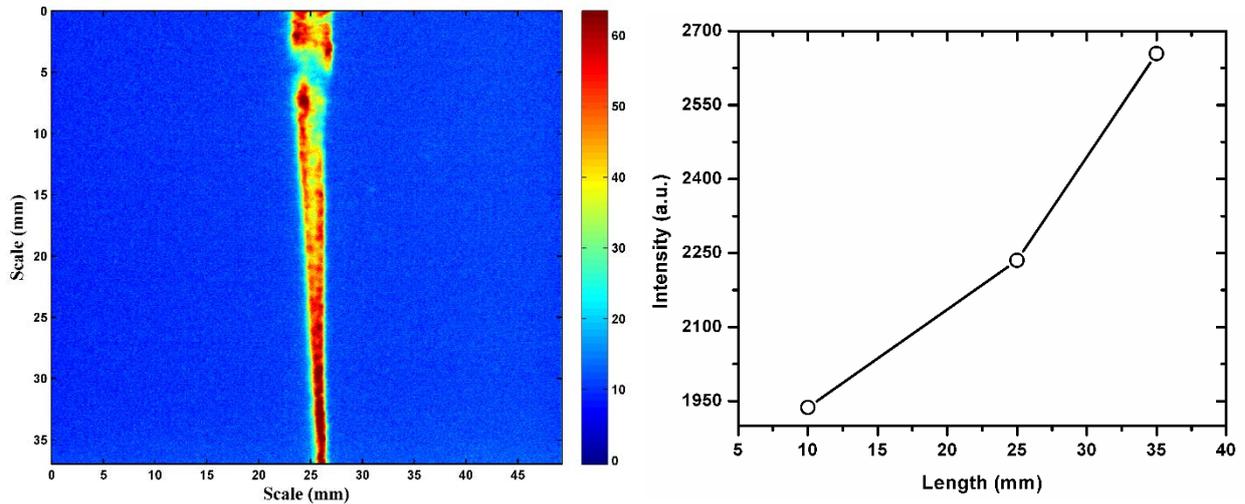

Figure 6 (colour online) (a) ICCD image of flowing plasma with 20 us exposure time. (b) Intensity profile of the plume over the length.

In summary, we have presented the estimation of plasma density along the length of the plume in the ambient air. By using a high voltage (4 kV$_{p-p}$ and 33 kHz frequency), and Helium gas of 11 lpm, a steady low-temperature atmospheric plasma plume of length around 4 cm has been produced in the ambient air. The real-time high-speed imaging of the plume captured by using the ICCD camera shows the plume expansion dynamics further explained by the drag force model shows the velocity measurements. Initially, the plume travels at 18.6 km/s, which rapidly slows down and reaches a final velocity of 2.1 km/s. Further, the measured plume current along the plasma plume length at 10, 25, and 35 mm is the range of (1.27-2) mA. By utilizing the above drift velocity and current information, the plasma density along the plume length has been estimated which is in the range of $(0.005\text{-}3.2) \times 10^{12}$ cm$^{-3}$. This increase in the density is due to the combined effect of the reduction of drift velocity, charge in the plume and area confinement in the plume. Further, it is validated by the images obtained with high exposure time. These above-mentioned characterizations of plasma plume have a very pertinent to a large number of biological and industrial applications.

**Acknowledgments:** Authors would like to thank Director, IPR, for providing funding and facilities to carry out this research work, Dr. Anitha V.P. for her support, Dr. Rajesh Kumar Singh for technical discussion.